\documentclass[twoside,12pt]{article}
\usepackage{epsfig,graphicx}

\newcommand{\beq}{\begin{equation}}
\newcommand{\eeq}{\end{equation}}
\newcommand\beqa{\begin{eqnarray}}
\newcommand\eeqa{\end{eqnarray}}
\newcommand\bea{\begin{array}}
\newcommand\eea{\end{array}}

\newcommand{\neqa}{\nonumber\end{eqnarray}}
\newcommand{\la}{\label}
\newcommand{\noi}{\noindent}
\newcommand{\eq}[1]{eq.(\ref{#1})}
\newcommand{\eqs}[2]{eqs.(\ref{#1},\ref{#2})}
\newcommand{\Eq}[1]{Eq.(\ref{#1})}

\newcommand{\ur}[1]{(\ref{#1})}

\newcommand{\re}{\relax{\rm I\kern-.18em R}}

\def\su2{{SU(2)}}

\begin{document}

\title{{\bf The narrow pentaquark}~\footnote{This is the write-up of the talks given at {\it Quarks-2006} (St.~Petersburg,
May 19-26, 2006) and at {\it Quark Confinement and Hadron Spectrum VII} (Ponta Delgada,
Sep. 2-7, 2006), to be published in the corresponding Proceedings.}}

\author{\bf Dmitri ~Diakonov
\\
\small{\em Petersburg Nuclear Physics Institute, Gatchina, 188 300,
St. Petersburg, Russia}
}
\maketitle

\begin{abstract}
The experimental status of the pentaquark searches is briefly reviewed.
Recent null results by the CLAS collaboration are commented, and new
strong evidence of a very narrow $\Theta^+$ resonance by the DIANA collaboration
is presented. On the theory side, I revisit the argument against
the existence of the pentaquark -- that of Callan and Klebanov -- and show
that actually a strong resonance is predicted in that approach, however its
width is grossly overestimated. A recent calculation gives 2 MeV for the
pentaquark width, and this number is probably still an upper bound.
\end{abstract}

\section{Experimental status}

The original claim for the discovery of a narrow exotic baryon
resonance in two independent experiments by T.~Nakano {\it et al.}~\cite{Nakano1}
and A.~Dolgolenko {\it et al.}~\cite{Dolgolenko1}, announced in the end of 2002~\footnote{They
were totally independent as both groups didn't know about the work of
one another and made a tedious re-analysis of data taken long before,
however both searches were triggered off by the authors of Ref.~\cite{DPP97}
where the resonance at $\sim 1530\,{\rm MeV}$ and width less than $15\,{\rm MeV}$
had been predicted.}, were followed in 2003-04 by a dozen experiments confirming
the resonance and about the same amount of non-sighting experiments. In 2005 the results
of the two CLAS high-statistics experiments were announced~\cite{CLAS-d2,CLAS-p2},
which didn't see a statistically significant signal of the $\Theta^+$ resonance
in the $\gamma d$ and $\gamma p$ reactions and gave upper bounds for its production
cross sections. Although those upper bounds didn't contradict the theoretical
estimates (see below) many people in the community jumped to the conclusion that
``pentaquarks do not exist".

Meanwhile, in 2005-06 new results became available~\cite{Nakano2,Dolgolenko2}
partly based on new data, confirming seeing the $\Theta^+$.

Usually, if one suspects a resonance in a system $A\!+\!B$, the best thing is
to study the {\it formation} of the resonance in $AB$ scattering, in this case in
the $K^+n$ (or $K^0p$) scattering. Unfortunately, mankind has lost $K^+$ and $K^0$
beams at the appropriate (low) energies, therefore most of the processes
studied so far are of the {\it production} type. We do not have much experience
with pentaquarks, which does not make it easy to estimate the production
cross sections to be able to judge that this or that non-observation of
a resonance ``kills" it.

\begin{minipage}{.50\textwidth}
\hspace{0.8cm}{\epsfig{figure=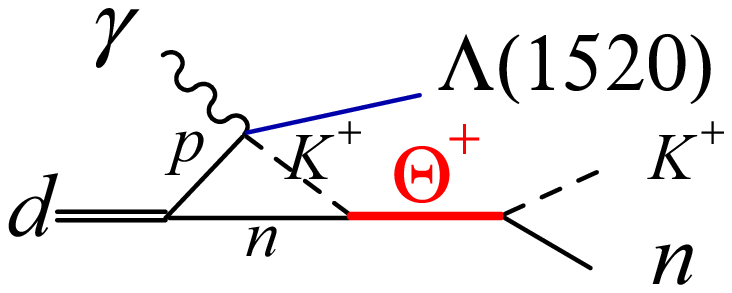,width=4cm}}
\vskip 0.5true cm

{\bf Fig.~1a,b}. Above: reaction studied by LEPS. Right: the spectrum of mass of the
$K^+n$ system~\cite{Nakano2}. The red histogram shows the estimated background. One observes
a narrow peak at 1.54 GeV.
\end{minipage}
\hskip 1.5true cm
\begin{minipage}{.50\textwidth}

{\epsfig{figure=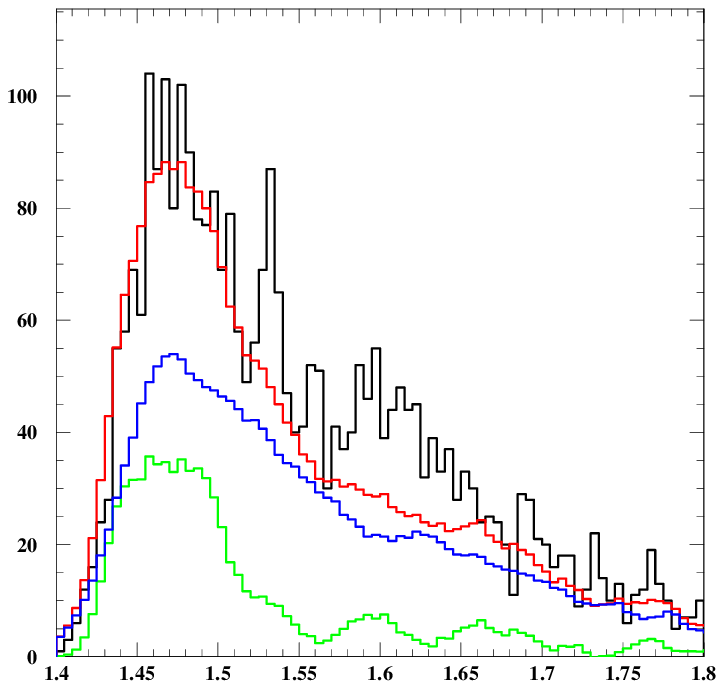,width=5.9cm}}
\end{minipage}
\vskip 0.3true cm

A {\it quasi-}formation experiment where a quasi-free $K^+$ scattered off
a quasi-free neutron inside a deuteron in $\gamma d$ reaction was
performed at SPring-8, near Osaka; the results were reported by T.~Nakano with
many details at a number of conferences~\cite{Nakano2}: there is a clear resonance
signal.

Why LEPS collaboration at SPring-8 sees the $\Theta^+$ peak whereas CLAS collaboration
at the Jefferson Lab does not? In both cases it is the same $\gamma d$ reaction at
comparable energies, however the kinematics and the detector acceptance are different.
Having a model for the reaction it is possible to compute the $K^+n$ mass spectrum
adjusted to the kinematical cuts imposed by the two collaborations and their apparatuses.
This has been done in Ref.~\cite{Titov}, see Fig.~2.

\begin{center}
{\epsfig{figure=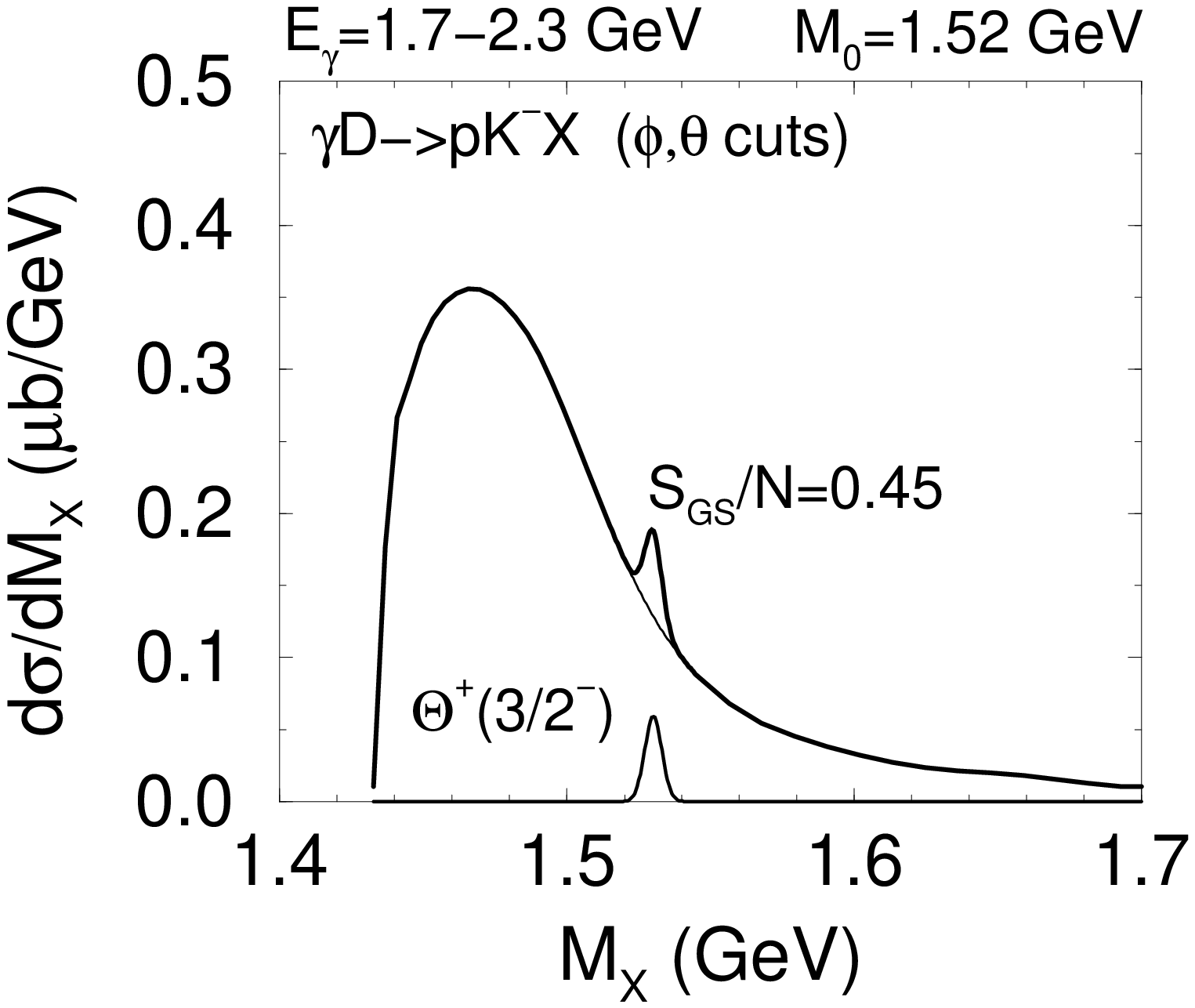,width=6cm}}\hskip 1true cm
{\epsfig{figure=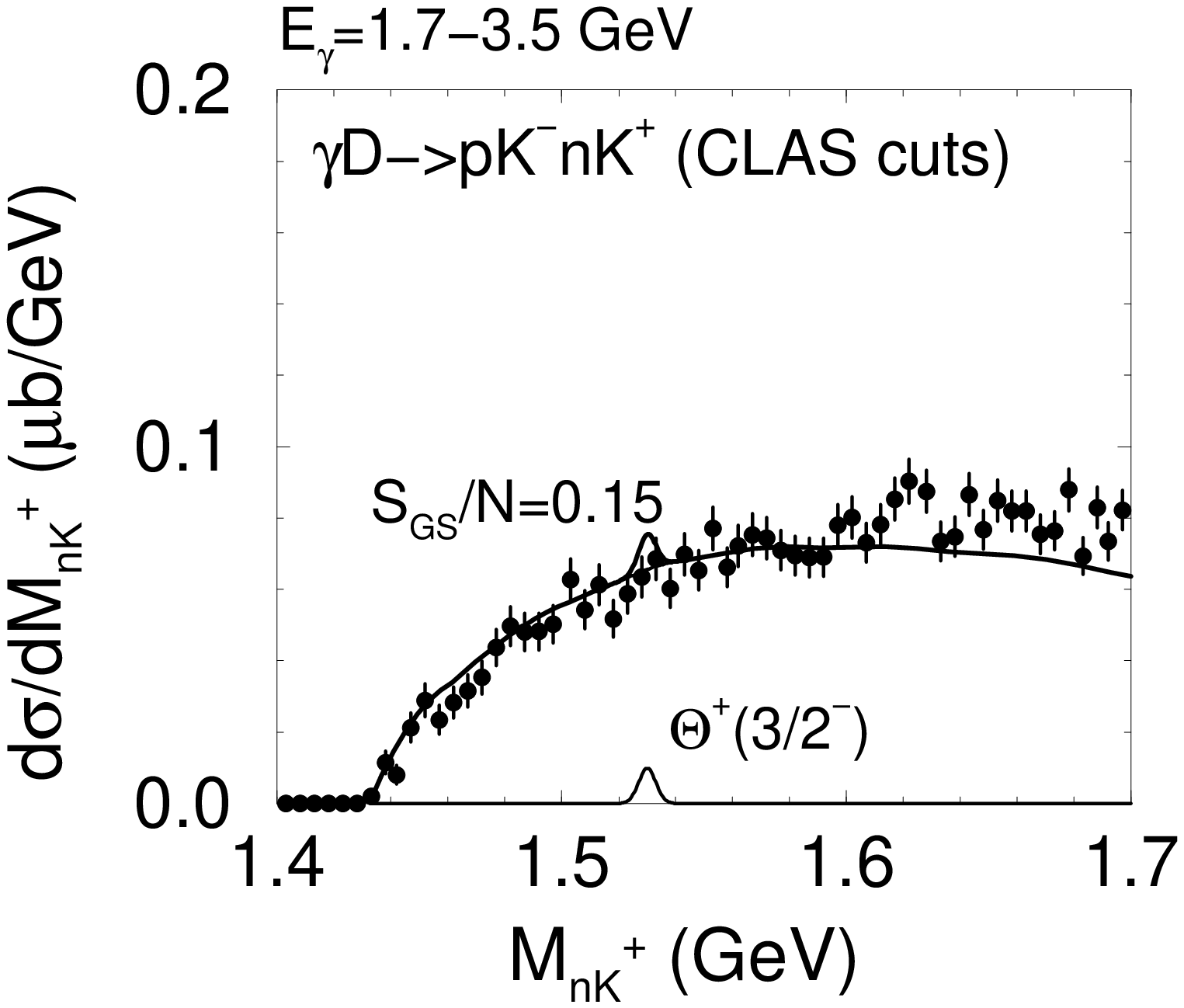,width=6cm}}
\end{center}

\noi {\bf Fig.~2 a,b} (from Ref.~\cite{Titov}. $\Theta^+$ should be mildly visible
in the LEPS setup (left) but buried under the background in the CLAS setup (right).
With most CLAS data points lying on the calculated solid curve, the authors of
Ref.~\cite{Titov} demonstrate a fair control of the background. Spin-parity $3/2^-$
has been assumed for the $\Theta^+$; were it $1/2^+$ the signal-to-background would
be worse.
\vskip 0.5true cm

A similar conclusion has been recently drawn by V.~Guzey~\cite{Guzey} from
evaluating the cross section of the process $\gamma d\to\Lambda K^+n$
also studied by the CLAS collaboration with no statistically significant
resonance structure observed. The claim is that in the CLAS setup~\cite{CLAS-d2Lambda}
the $\Theta^+$ signal would be almost completely washed out through
interference with non-resonant processes.

CLAS collaboration studied also the $\Theta^+$ production in the
$\gamma p\to\bar K^0(K^+n)\,({\rm and}\;K^0p)$ reaction~\cite{CLAS-p2}, again with
a null result. It should be noted that it is different from the
$\gamma p\to\pi^+K^-K^+n$ reaction where a $7\sigma$ signal of the $\Theta^+$
has been previously reported by the same collaboration~\cite{CLAS-p1}.
The reaction $\gamma p\to \bar K^0\Theta^+$ is a simple 2-body one, and its
cross section can be estimated more or less reliably from the (reggeized)
vector $K^*$ exchange. The vector $K^*$ coupling to the $p\Theta$ transition
vanishes in the $SU(3)$ limit, so it couples through the magnetic moment
vertex, $\sigma_{\mu\nu}q_\nu$, but even this coupling is expected to be
an order of magnitude less than for the octet-octet and octet-decuplet
magnetic transitions~\cite{Polyakov-magnetic}. The estimate of the $\gamma p\to
\bar K\Theta$ yield has been made in Ref.~\cite{K*exchange} prior
to the CLAS experiment, with the result of about 0.2 nb. The CLAS experimental
upper limit of $\sim 0.7$ nb for the $\Theta$ production~\cite{CLAS-p2} is,
therefore, not too restrictive. One can hardly conclude from these numbers
that ``$\Theta^+$ does not exist". However, the impressive amount of data
collected by CLAS allows one to hope that a clever analysis combined with
reliable theoretical estimates may really bury (or reveal) the $\Theta^+$.

I do not discuss here the numerous non-sighting experiments at high energies:
the exotic baryon production cross sections are not known there. It can be argued,
however, that the $\Theta^+$ production at high energies is at least an order of
magnitude less than that of the $\phi$ meson and two orders of magnitude less than
of the $\Lambda$ hyperon~\cite{D04}. However those ratios may vary depending
on the concrete experimental setup.

Finally, let me draw attention to the {\em direct formation}
experiment~\cite{Dolgolenko2} which, in my mind, gives to date the most strong evidence
in favour of a very narrow $\Theta^+$. It is the DIANA experiment at ITEP, Moscow,
-- actually the update of their first analysis of the $K^+n({\rm Xe})\to K^0p$
data~\cite{Dolgolenko1} but now with approximately double statistics. Previously,
there were about 30 events above the estimated background, now there about 60,
as it should be if the signal is real. Also, more thorough analysis has been performed
to understand better the kinematics of the reaction and the background processes,
see Fig.~3. The resonance peak is seen already in the raw data (white histogram)
but is strongly enhanced by a mild kinematical cut suppressing re-scattering (grey
histogram).

\hskip 2true cm {\epsfig{figure=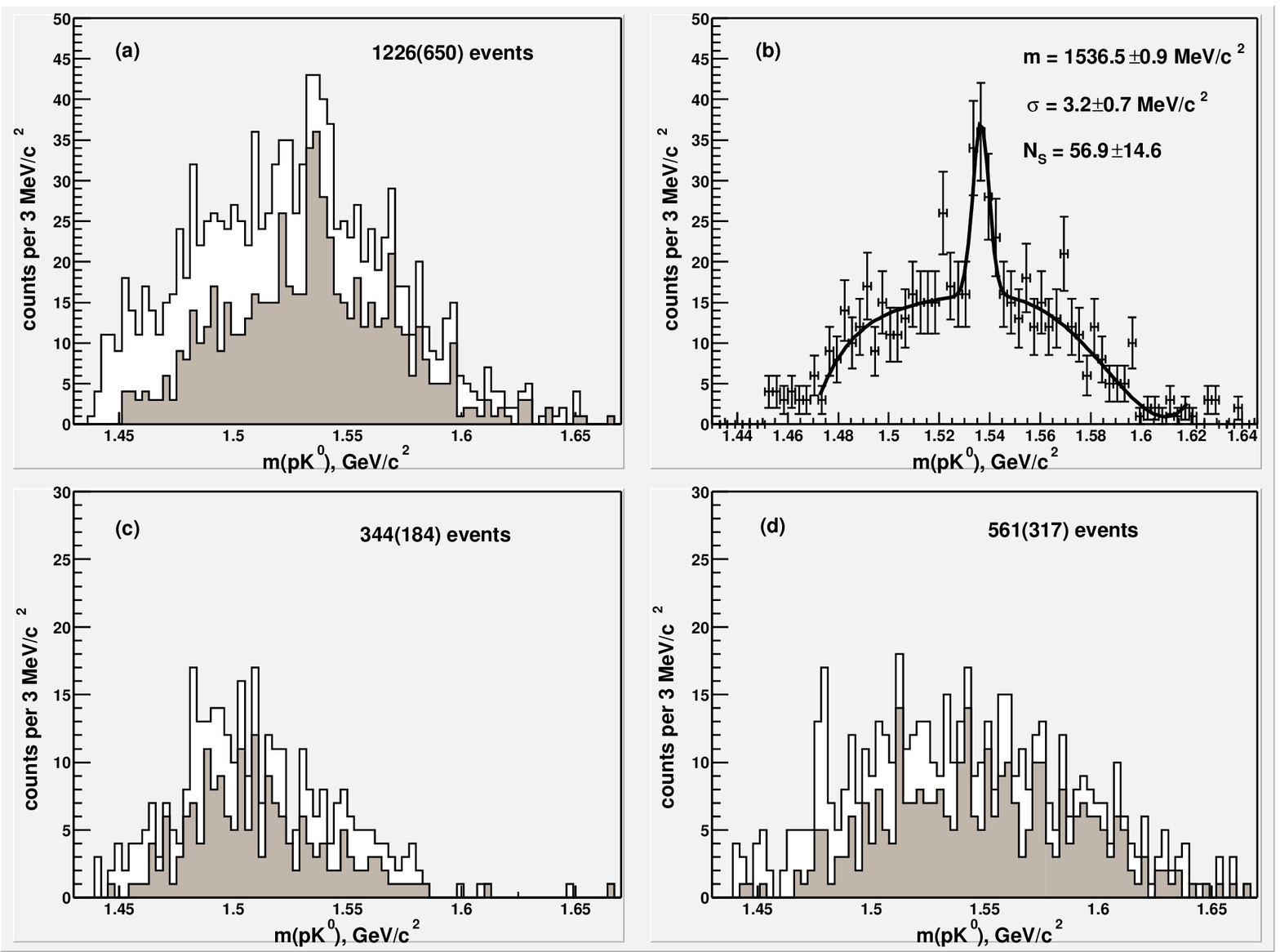,width=12.5cm}}

\noi
{\bf Fig.~3} (from Ref.~\cite{Dolgolenko2}). $K^0p$ mass distribution
in the range of the incident $K^+$ momenta where the $\Theta^+$ resonance {\em can be}
formed due to neutron's Fermi motion ({\bf a,b}), and where it {\em cannot be}
formed: $K^+$ momenta are either too low ({\bf c}) or too high ({\bf d}).
In {\bf c,d} the invariant mass spectrum is filled owing partly to re-scattering
of final particles in the Xe nucleus, cf.~\cite{Sibirtsev}.
\vskip 0.2true cm

If this is not a resonance, what is it? The authors find its statistical significance
to be 7.3$\sigma$, 5.3$\sigma$, 4.3$\sigma$, depending on
whether one estimates it as $S/\sqrt{B}$, $S/\sqrt{S+B}$ or $S/\sqrt{S+2B}$.
The mass is found to be $M_{\Theta}=1537\pm 2\,{\rm MeV}$ and the width
$\Gamma_{\Theta}=0.36\pm 0.11\,{\rm MeV}$ (!) (plus possible systematic uncertainties).
This is the only experiment where the direct estimate of the width is possible
since the formation cross section averaged over the resonance range is proportional
to the width. The only other available formation experiment with the secondary kaon
beam at BELLE sets an upper limit of $\Gamma_\Theta$ beyond the above value~\cite{BELLE}.

We have to keep in mind that there are numerous and so far uncontested observations of a
$K\!N$ resonance at 1.53 GeV in neutrino-~\cite{neutrino}, photon-~\cite{photon} and
proton-~\cite{proton} induced reactions. The analysis of old $K^+d$ data by Gibbs
calls for the exotic resonance with the width $0.9\pm 0.3\,{\rm MeV}$. An anomaly in
$K^+$ scattering off nuclei needs an ``additional reactivity'' as compared to
the usual optical potential scattering~\cite{Gal}. Last but not least, the GRAAL
collaboration reports a possible narrow $N^*(1675)$ resonance in the $\gamma n\to \eta n$
reaction (but not in the $\gamma p\to\eta p$)~\cite{Kuznetsov} which is consistent
with the resonance being the antidecuplet partner of the $\Theta^+$.

Given a small $KN\Theta$ coupling constant (since the width is very small)
and a small $K^*N\Theta$ coupling (since the transition magnetic moment is small),
it is difficult to arrange for a sizable production of the $\Theta^+$. Maybe a good
chance of seing it is via an interference with some process with large amplitude --
then at least the cross section is proportional to the small coupling but not its square.
However, if $\Theta^+$ is produced through interference, it becomes hostage of the
specific conditions of a reaction: the resonance may appear as a peak or a dip
or an oscillation, depending on the relative phase of the amplitudes. One may be lucky
in one setup and less lucky in another, as Guzey's example~\cite{Guzey} has shown.

The direct formation experiment~\cite{Dolgolenko2} reveals $\Theta^+$
and the quasi-formation experiment~\cite{Nakano2} sees it, too. The high-statistics
CLAS $\gamma d$ and $\gamma p$ experiments impose upper limits on the production
cross sections, which seem so far to be beyond the danger zone for the $\Theta^+$.
The high energy probes also impose certain limits on the production, but at present
it is not easy to translate them into physical meaning.

Future progress can be obtained along the following lines: a) by performing a high-flux
$KN$ formation experiment (planned at J-PARC), b) by learning to make reliable
estimates for the production cross sections, such that the comparison with the
data becomes meaningful, and c) by inventing clever new methods of searching
$\Theta^+$ taking into account that all its couplings to normal hadrons are small.

\section{Theoretical surprise}

Probably the only theoretical argument against the existence of exotic baryons
is due to Callan and Klebanov~\cite{CK}~\footnote{T.~Cohen gave an additional
argument~\cite{Cohen} why the Callan--Klebanov approach to the exotic baryon
must be correct at large number of colours.}. It relies on the academic
limit of large number of colours $N_c$ when baryons can be considered in
the mean field approximation with quarks bound by the self-consistent
pion field, the ``soliton'' ({\it \`a la} large-Z Thomas--Fermi atom or
the large-A shell model for nuclei). The Skyrme model is a popular realization
of this idea, although not a too realistic one~\cite{DP00}.

In this approach, octet and decuplet baryons are all rotational excitations --
in ordinary and flavour spaces -- of the same object, the large `classical' baryon.
At large $N_c$, however, baryons with minimal strangeness (like the nucleon,
the $\Delta$, the $\Theta$...) correspond to rotational states which are more
like a precession along a high latitude around the ``North pole''~\cite{Cohen,DP05}.
Such a rotation can be as well considered as a small oscillation about the
pole. Therefore, at large $N_c$ the existence or non-existence of the
$\Theta^+$ can be studied by considering small oscillations of the kaon field
about the `classical' nucleon, which I shall generically call the `Skyrmion'.
In other words, it is sufficient to look into the kaon scattering off a
Skyrmion, and that is what Callan and Klebanov did. After the discovery of the
$\Theta^+$, the study has been repeated in more detail in Ref.~\cite{Klebanov}.
One has to solve a Schr\"odinger-Klein-Fock equation~\footnote{Historical survey by
Jackson and Okun~\cite{JO} disclosed that the relativistic wave equation for
spin 0 particle has been first written down and published in {\it Zeitschrift
f\"ur Physik} in 1926 practically simultaneously by E.~Schr\"odinger, O.~Klein
and V.~Fock. Moreover, Fock's is {\em the} paper where gauge invariance
in quantum theory was first introduced (he called it the gradient invariance).
Gordon's paper on the application of the already known equation came later.
Therefore, I do not see reasons to prolong historical injustice, and call the
relativistic wave equation by its proper name.}
but with a Wess--Zumino--Witten term linear in the time derivative, and
find the scattering phases for given quantum numbers. The resulting phase
in the strangeness $+\!1$, spin $1/2^+$ channel is plotted in Fig.~4a.
\vskip 0.3true cm

{\epsfig{figure=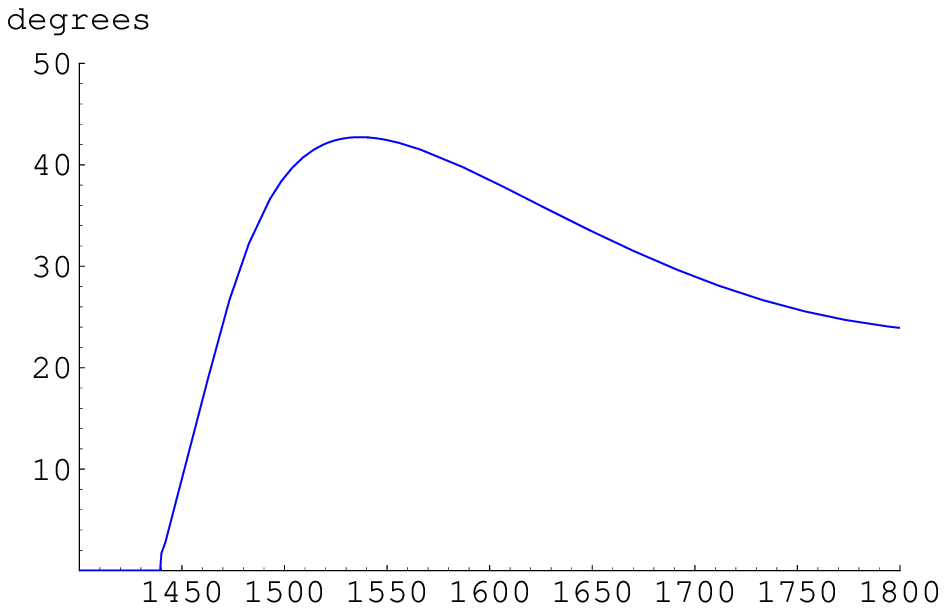,width=6.5cm}}
\hskip 2true cm{\epsfig{figure=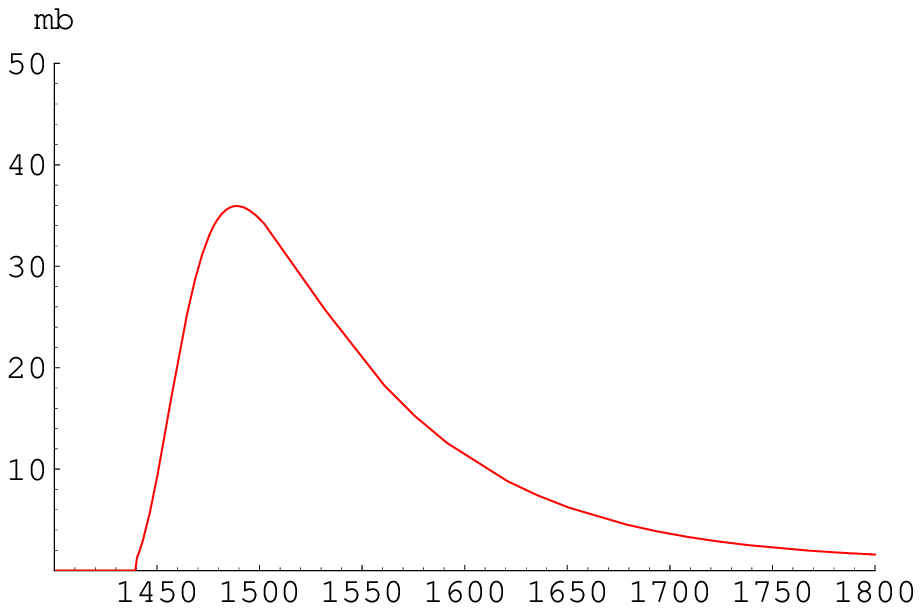,width=6.5cm}}
\vskip 0.3true cm

\noi {\bf Fig.~4a,b}. The $K^+n$ scattering phase~\cite{Klebanov} (left)
and the ensuing $K^+n$ scattering cross section (right) as function of
the invariant $K^+n$ mass in the Skyrme model. Note that at the maximum the
cross section is as large as 35 mb ! Courtesy V.~Petrov.
\vskip 0.3true cm

The point made in Refs.~\cite{CK,Klebanov} is that the $K^+n$ phase shift in Fig.~4a
does not pass through $90^{\small o}$ as it should be for an isolated
Breit--Wigner resonance, and therefore there is no exotic resonance, at least
in the large-$N_c$ limit. However, if there is both a resonance {\em and} a
potential scattering, the phase shift needs not go through $\pi/2$.

To see what is going on, it is instructive to solve the Callan--Klebanov $K^+n$
scattering equation in the complex energy plane, simultaneously varying the coefficient
in front of the Wess--Zumino--Witten term~\cite{Petrov}. When it is zero, there
is exact zero-energy solution corresponding to the rotation of the soliton as
a whole in the flavour space. It was on the base of the quantization of this rotation
that the light and narrow $\Theta^+$ was predicted~\cite{DPP97}. As one increases the
coefficient of the Wess--Zumino--Witten term towards its physical value, the
would-be zero energy level moves up but obtains an imaginary part. With the standard
Skyrme model parameters used by Klebanov {\it et al.}, the pole position of
the $\Theta^+$ resonance is at $E_{\Theta}=1510-\frac{i}{2}\!\cdot\!120\,{\rm MeV}$. Indeed,
had Klebanov {\it et al.}~\cite{Klebanov} plotted the $K^+n$ cross section from their phase
shift according to the well-known formula $\sigma=(4\pi/k^2)(2j+1)\sin^2\delta$,
they would get a very strong resonance, see Fig.~4b.

Thus, the prediction of the Skyrme model is not that there is no exotic resonance
but just the opposite: {\bf there is a very strong resonance}, at least
when the number of colours is taken to infinity! Therefore, a theorist must be worried
not by the existence of an exotic resonance but rather by its absence: why a very general
theoretical prediction -- a broad exotic resonance -- is not observed in
nature~\footnote{Varying the parameters of the Skyrme model~\cite{Klebanov,WW}
or modifying it~\cite{Rho} can make the exotic resonance narrower or broader
but one cannot get rid of it. The reason is very general: energy levels do not disappear
as one varies the parameters but move into the complex plane.}.

The answer is that the Callan--Klebanov large-$N_c$ logic in general and the concrete Skyrme
model in particular grossly overestimate the resonance width. In reality it becomes very narrow,
and that is why it is so difficult to observe it. We first deal with the large-$N_c$
limit and check if it is a good approximation for the $\Theta^+$ resonance. A general argument
has been presented in Ref.~\cite{DP05} that it is not but here we give a more direct argument.

Let us recall the equation for the $\Theta^+$ width~\cite{DPP97}~\footnote{The kinematical
factor in \eq{1} is written in the non-relativistic limit for simplicity; its precise form
is irrelevant for the present discussion.}:
\beq
\Gamma_\Theta=\frac{3|{\bf p}|^3}{2\pi(M_N+M_\Theta)^2}
\cdot\frac{1}{5}\cdot\left(G_0-G_1-\frac{1}{2}\,G_2\right)^2
\la{1}\eeq
where ${\bf p}$ is the 3-momentum of the kaon and $G_{0,1,2}$ are axial
couplings appearing as constants in front of different $SU(3)$ structures.
\Eq{1} is written for the real world, $N_c=3$, however Michal Praszalowicz has
generalized it to the world with arbitrary $N_c$~\cite{MichalNc}:
\beq
\Gamma_\Theta=\frac{3|{\bf p}|^3}{2\pi(M_N+M_\Theta)^2}
\cdot\frac{3(N_c+1)}{(N_c+3)(N_c+7)}\cdot\left(G_0-\frac{N_c+1}{4}\,G_1-\frac{1}{2}\,G_2\right)^2.
\la{2}\eeq
To join the Callan--Klebanov logic, one has first to take the limit $N_c\!\to\!\infty$
(because only in this limit one can replace large-angle rotation in the flavour space by small
oscillations of the kaon field) but then put $N_c\!=\!3$ in the final result in
order to compare it with \eq{1} written at $N_c\!=\!3$ from the start. One should mind that
$G_0\!=\!{\cal O}(N_c^{\frac{3}{2}}),\,G_{1,2}\!=\!{\cal
O}(N_c^{\frac{1}{2}}),\,M_N\!=\!{\cal O}(N_c)\!=\!M_{\Theta},\,|{\bf p}|\!=\!{\cal O}(1),\,
\Gamma_\Theta={\cal O}(1)$.
This operation leads to
\beq
\Gamma_\Theta^{\rm CK}=\frac{3|{\bf p}|^3}{2\pi(M_N+M_\Theta)^2}
\cdot 1\cdot\left(G_0-\frac{3}{4}\,G_1\right)^2.
\la{3}\eeq
This width is expected to correspond to the imaginary part of the pole position
in the Callan--Klebanov scattering problem. Comparing \eqs{1}{3} one clearly sees
what happens when the limit $N_c\!\!\to\!\!\infty$ is used: the width is first
increased by a factor of 5 (!) and then may be further increased by a more shallow cancelation of the
constants $G_{0,1,2}$. These constants as well as the masses have also $1/N_c$
corrections but those are expected to be additionally suppressed by powers of
$1/2\pi$, see below.

The conclusion is that the exotic resonance is theoretically inevitable but
that its small width cannot be obtained in the large-$N_c$ limit. Had $N_c$
been 300 instead of 3, $\Theta^+$ could be as broad as any other well-established
baryon resonance. It would have been produced in abundance in hadron collisions.

\section{Estimate of the $\Theta^+$ width}

Forbidding ourselves to use large $N_c$ as a theoretical tool we get in trouble.
However, one can still use the Relativistic Mean Field Approximation (RMFA)~\cite{DP05}
(alias the Chiral Quark Soliton Model~\cite{DP00}). Being a relativistic field-theoretic
model, it allows to account for quark pair creation and annihilation in a consistent way,
and that is what we need here.

The RMFA is generally justified when $N_c$ is large. At a closer look, however,
one can see that there are two types of $1/N_c$ corrections to the mean-field
results. One type comes from high-frequency fluctuations; these are in fact
meson loop corrections that bring in additional powers of $1/2\pi$. These
corrections go in powers of $1/(2\pi N_c)\approx 6\%$ and will be ignored
at the present level of accuracy. I remind that in QED the actual expansion
parameter from radiative corrections is not $\alpha=1/137$ but rather
$\alpha/2\pi \sim 10^{-3}$. The success of Wilson's $\epsilon$-expansion
in computing anomalous dimensions for critical phenomena is due to the fact
that the actual expansion parameter is not $\epsilon=1$ but rather $\epsilon/2\pi$.

Other type of corrections to the mean field arise from low frequencies and are
all related to zero modes, {\it viz.} translations and rotations of the ``soliton''
as a whole. These corrections are ${\cal O}(1/N_c)$ but are not accompanied by
additional small factors $1/2\pi$. An example of such correction is presented by
the Clebsch--Gordan coefficient in \eq{2}:
\beq
\frac{3(N_c+1)}{(N_c+3)(N_c+7)}=\frac{3}{N_c}\left(1-\frac{9}{N_c}+\frac{69}{N_c^2}
-\frac{501}{N_c^3}+\ldots\right)
\la{4}\eeq
Apparently one cannot trust the result of the leading order when $N_c\!=\!3$.
Such corrections must be summed up exactly, which is equivalent to treating the
rotations exactly at the physical value $N_c\!=\!3$.\\
\vskip 0.5true cm

In the mean field approximation {\em all} quark wave functions inside {\em all}
baryons belonging to the octet, decuplet and exotic antidecuplet are known for
{\em all} their Fock, i.e. $3Q,\,5Q,\,7Q,...$ components~\cite{DP05}. The leading
component in the ordinary octet and decuplet baryons is naturally the $3Q$ one
(judging from its normalization) but there is a sizable ($\sim$30\%) addition of
the $5Q$ component. For some baryon observables the $5Q$ component gives a mild
correction (and that is why the primitive $3Q$ constituent quark models are
not so bad as one would naively expect) but in some other observables higher
components are critical to obtain agreement with the experiment, {\it e.g.} to
explain the ``spin crisis'' or the large value of the nucleon $\sigma$-term.
About 30\% of the time nucleons are pentaquarks!

As to the exotic $\Theta^+$ and other members of the antidecuplet, their
{\em lowest} Fock component is the $5Q$ one, nothing terrible. The spatial wave function
of 5 quarks in $\Theta^+$ is very similar to that of the $5Q$ component of the
nucleon, only the spin-flavour part of the wave function is somewhat different.
The extra $Q\bar Q$ pair in the $\Theta^+$ is a (known~\cite{DP05}) mixture of
$0^+,0^-,1^-$ and $1^+$ waves corresponding to scalar, pseudoscalar, vector and
axial mesons. However, they do not form `molecules' as the `mesons' are deep
inside the `$3Q$ baryons' in pentaquarks.

To evaluate the width of the $\Theta^+\to K^+n$ decay one has to compute the
transition matrix element of the strange axial current,
$<\!\Theta^+|\bar s\gamma_\mu\gamma_5u|n\!>$. The important point is that there
are, generally, two contributions to this matrix element: the ``fall apart''
process (Fig.~5, A) and the ``5-to-5'' process where $\Theta^+$ decays into
the $5Q$ component of the nucleon (Fig.~5, B). I stress that one does not exist
without the other: if there is a ``fall apart'' process it means that there
is a non-zero coupling of quarks to pseudoscalar (and other) mesons, meaning
that there is a transition term in the Hamiltonian between $3Q$ and $5Q$ states
(Fig.~5, C). Hence the eigenstates of the Hamiltonian must be a mixture of
$3Q,5Q,...$ Fock components. Therefore, assuming there is process A, we have to
admit that there is process B as well.
\vskip 0.5true cm

\centerline{\epsfxsize=12cm\epsfbox{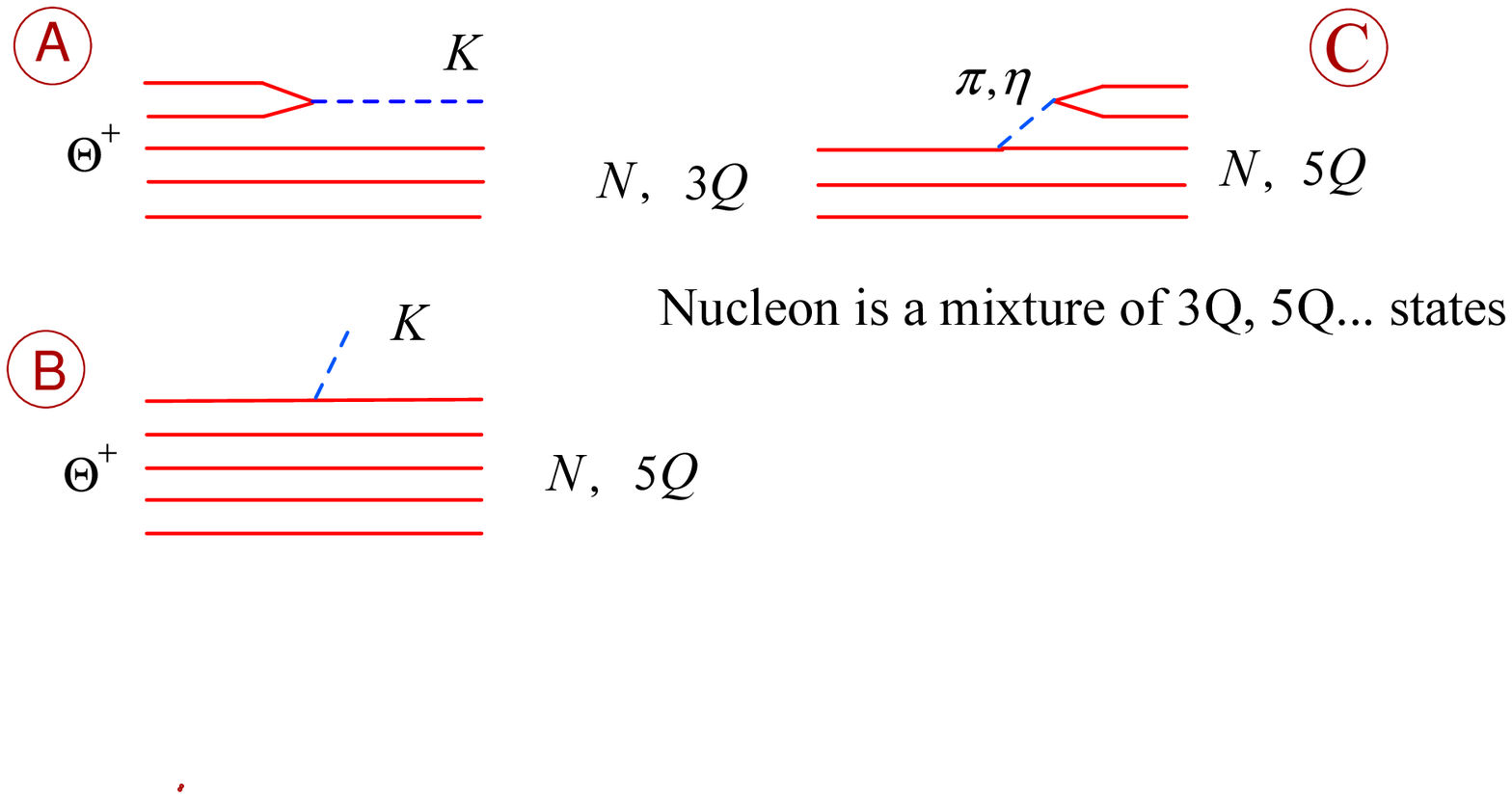}}
\noi
\vskip -1.true cm

\noi {\bf Fig.~5}. Contributions A and B to the $\Theta^+\to K^+n$ decay.
\vskip 0.5true cm

Moreover, each of the amplitudes A and B are not Lorentz-invariant, only
their sum is. Evaluating the ``fall-apart'' amplitude and forgetting about
the ``5-to-5'' one makes no sense. For example, in the lab frame there is a
tendency for the two amplitudes to cancel each other (A.~Hosaka, private
communication). A convenient way to evaluate the sum of two graphs, A and B,
is to go to the infinite momentum frame (IMF) where only the process B
survives, as axial (and vector) currents with a finite momentum transfer do not
create or annihilate quarks with infinite momenta. The baryon matrix elements
are thus non-zero only between Fock components with {\em equal number} of
quarks and antiquarks. We note that in the RMFA, moving from one frame to
another just requires a Lorentz transformation of the mean field
and of the corresponding vector and spinor fields without changing the form
of the mean field. This is seen, {\it e.g.}, from comparing two different ways of
calculating nucleon parton distributions in the RMFA, leading to the same
results~\cite{DIS}.

The transition matrix element of the strange axial charge,
$<\!\Theta^+|\bar s\gamma_0\gamma_5u|n\!>$ was evaluated in the IMF in
Ref.~\cite{DP05} with the resulting width $\Gamma_\Theta=2\,{\rm to}\,4\,
{\rm MeV}$. The uncertainty was mainly due to the uncalculated quark-exchange
contributions and relativistic corrections. These were subsequently computed
by C\'edric Lorc\'e~\cite{Lorce} with the result

\beq
\Gamma_\Theta\approx 2\,{\rm MeV}.
\la{5}\eeq

\noi It should be noted that the calculation of the above matrix element was performed
assuming the chiral limit for the kaon and zero momentum transfer. In fact the
momentum transfer in the $\Theta^+\to K^+n$ decay is several hundred MeV,
therefore one must expect a further formfactor-type suppression of the estimate
\ur{5} such that $\Gamma_\Theta$ may well end up at the sub-MeV level which is
where the current value of the width is~\cite{Dolgolenko2}.

The physical reason why the axial constant for the $\Theta\to N$ transition
$g_{\Theta NK}\approx 0.14$ appears to be an order of magnitude less than the
nucleon constant $g_{N}=1.26$ (resulting in the suppression of the $\Theta$ width
by two orders of magnitude as compared to the normal 100 MeV width for strongly decaying
baryons) is clearly seen from the calculations~\cite{DP05,Lorce}. The large value
of the axial constant in normal baryons in mainly due to their $3Q$ component,
the $5Q$ component contributing much less. However, it is the latter contribution
that is comparable to the axial constant $g_{\Theta NK}$ as it is a $5Q$
effect, too. It is suppressed to the same extent as is the $5Q$ component in ordinary
baryons. As stressed in our first publication~\cite{DPP97}, in the imaginary
non-relativistic limit when ordinary baryons are made of three quarks with no admixture
of $Q\bar Q$ pairs the $\Theta^+$ width tends to zero strictly. \\

To summarize: The very small width of $\Theta^+$ is natural; the present
estimate \ur{5} will probably go down when formfactor suppression is included.
We have revisited the theoretical argument of Callan and Klebanov against the
exotics and found that actually it is the opposite: the Skyrme model at large
$N_c$ predicts a too strong resonance. We have shown, however, that a broad
width is a very-large-$N_c$ artifact. On the experimental side, there is new
strong evidence of an extremely narrow $\Theta^+$ from DIANA, a very significant
new evidence from LEPS, and other older evidence which is difficult to brush aside.
The null results from the new round of CLAS experiments are compatible with
what one should expect based on the estimates of production cross sections. \\

I thank A.~Dolgolenko, V.~Guzey, A.~Hosaka, T.~Nakano, A.~Titov and especially V.~Petrov
and M.~Polyakov for discussions. The paper has been written during the visit to Bochum
University, sponsored by the A.v.Humboldt Award. This work is supported
in part by Russian Government grants 1124.2003.2 and RFBR 06-02-16786.

\end{document}